\documentclass{ws-procs975x65}

\begin{document}

\title{STUDIES ON THE UV TO IR EVOLUTION OF GAUGE THEORIES AND 
QUASICONFORMAL BEHAVIOR}

\author{R. SHROCK$^*$}

\address{C. N. Yang Institute for Theoretical Physics, Stony Brook University,
 \\ Stony Brook, NY 11794, USA \\ and 
Department of Physics, Sloane Laboratory \\
Yale University, New Haven, CT 06520, USA 
$^*$E-mail: robert.shrock@stonybrook.edu}

\begin{abstract}

We describe recent results from our studies of the UV to IR evolution of
asymptotically free vectorial gauge theories and quasiconformal behavior.
These include higher-loop calculations of the IR zero of the beta function and
of the anomalous dimension of the fermion bilinear.  Effects of
scheme-dependence of higher-loop results are assessed in detail. Applications 
to models with dynamical electroweak symmetry breaking are discussed. 

\end{abstract} 

\keywords{UV to IR evolution, IR fixed point, quasiconformal behavior}

\bodymatter

\section{Introduction}
\label{intro}

The evolution of an asymptotically free gauge theory from the ultraviolet (UV)
to the infrared (IR) is of fundamental field-theoretic interest.  In this
SCGT12 talk we report on our calculations in \cite{bvh}-\cite{sch} with
T. A. Ryttov of higher-loop corrections to this UV to IR evolution and on some
new results (submitted shortly after the SCGT12 workshop in \cite{bc,lnn}).
The dependence of the gauge coupling $g(\mu)$ on the Euclidean momentum scale,
$\mu$, is determined by the $\beta$ function $\beta \equiv \beta_\alpha \equiv
d\alpha/dt$, where $t=\ln \mu$ and $\alpha(\mu)=g(\mu)^2/(4\pi)$ (and $\mu$
will often be suppressed in the notation).  This function has the series
expansion
\begin{equation}
\beta = -2\alpha \sum_{\ell=1}^\infty b_\ell \, a^\ell
         = -2\alpha \sum_{\ell=1}^\infty \bar b_\ell \, \alpha^\ell \ ,
\label{beta}
\end{equation}
where $a=\alpha/(4\pi)$, $\ell$ denotes the number of loops involved in the 
calculation of the coefficient $b_\ell$, and $\bar b_\ell=b_\ell/(4\pi)^\ell$.
The first two coefficients in $\beta$, $b_1$ and $b_2$, are scheme-independent
and were calculated in \cite{b1,b2}.  The $b_\ell$ have been calculated up to
4-loop order in the $\overline{MS}$ scheme in \cite{b3,b4}.  The
$n$-loop ($n \ell$) beta function, denoted $\beta_{n\ell}$, is given by the RHS
of Eq. (\ref{beta}) with the upper bound on the sum set equal to $n$ instead of
$\infty$.

We consider the UV to IR evolution of an asymptotically free vectorial
gauge theory with gauge group $G$ and $N_f$ massless fermions transforming
according to a representation $R$ of $G$. There are two possibiities for this
evolution: (i) there may not be any IR zero in $\beta$, so that as $\mu$
decreases, $\alpha(\mu)$ increases, eventually beyond the perturbatively
calculable region (which is the case for QCD); (ii) $\beta$ may have an IR zero
at a certain value denoted $\alpha_{IR}$, so that as $\mu$ decreases,
$\alpha(\mu)$ increases from 0 toward $\alpha_{IR}$. In this class of theories,
there are two further generic possibilities, namely $\alpha_{IR} < \alpha_{cr}$
or $\alpha_{IR} > \alpha_{cr}$, where $\alpha_{cr}$ is the critical minimal
value of $\alpha$ (depending on $R$) for spontaneous chiral symmetry breaking
(S$\chi$SB).  If $\alpha_{IR} < \alpha_{cr}$, then the zero of $\beta$ at
$\alpha_{IR}$ is an exact IR fixed point (IRFP) of the renormalization group;
as $\mu \to 0$ and $\alpha \to \alpha_{IR}$, $\beta \to \beta(\alpha_{IR})=0$,
and the theory becomes exactly scale-invariant with nontrivial anomalous
dimensions \cite{b2,bz}.

If $\beta$ has no IR zero, or an IR zero at $\alpha_{IR} > \alpha_{cr}$, then
as $\mu$ decreases through a scale denoted $\Lambda$, $\alpha(\mu)$ exceeds
$\alpha_{cr}$ and spontaneous chiral symmetry breaking occurs, so that the
fermions gain dynamical masses $\sim \Lambda$. In this case, in the low-energy
effective field theory applicable for $\mu < \Lambda$, one integrates these
fermions out, and the $\beta$ function becomes that of a pure gauge theory,
which has no (perturbative) IR zero. Hence, if $\beta$ has a zero at
$\alpha_{IR} > \alpha_{cr}$, this is only an approximate IRFP. If $\alpha_{IR}
> \alpha_{cr}$, the effect of the approximate IRFP at $\alpha_{IR}$ depends on
how close it is to $\alpha_{cr}$.  The desire to understand better both quantum
chromodynamics (QCD) and the properties of this IR zero have motivated
calculations of higher-loop terms in $\beta$ \cite{b3,b4} and higher-loop
corrections to the 2-loop result for the IR zero \cite{gkgg,bvh,ps}.  We denote
the IR zero of the $n$-loop beta function, $\beta_{n\ell}$, as
$\alpha_{IR,n\ell}$. The need to go beyond the 2-loop result for $\alpha_{IR}$
in studies of quasiconformal theories is evident from the fact that
$\alpha_{cr} \sim O(1)$ and one is interested in values of $\alpha_{IR}$ near
to $\alpha_{cr}$. Although the coefficients in $\beta$ at $\ell \ge 3$ loop
order are scheme-dependent, the results give a measure of accuracy of the
2-loop calculation of the IR zero, and similarly with the value of the
anomalous dimension of the fermion bilinear (discussed below).

If $\alpha_{IR}$ is only slightly greater than $\alpha_{cr}$, then, as
$\alpha(\mu)$ approaches $\alpha_{IR}$, since $\beta = d\alpha/dt \to 0$,
$\alpha(\mu)$ varies very slowly as a function of the scale $\mu$, i.e., there
is approximately scale-invariant (equivalently, dilatation-invariant,
slow-running, or ``walking'') behavior \cite{wtc}. For these theories, this is
equivalent to quasiconformal behavior \cite{fgs}. The spontaneous chiral
symmetry breaking and attendant fermion mass generation at $\Lambda$
spontaneously break the approximate dilatation symmetry, plausibly leading to a
resultant light Nambu-Goldstone boson, the dilaton \cite{dil}. The dilaton is
not massless, because $\beta$ is not exactly zero for $\alpha(\mu) \ne
\alpha_{IR}$. Studies of gauge-singlet bound states in confining quasiconformal
gauge theories, using Schwinger-Dyson and Bethe-Salpeter equations, find a
significant reduction in the scalar mass divided by the vector meson mass
\cite{sg}. Eventually, lattice gauge measurements may be able to determine the
mass of a dilaton in a theory of this type.  The quasiconformal behavior
associated with an approximate IRFP has been an ingredient of DEWSB models
since the 1980s, as a means of enhancing Standard-Model (SM) fermion masses
while keeping neutral flavor-changing current (FCNC) processes sufficiently
suppressed \cite{wtc}.

\section{Basic Properties of $\beta$}

Since $b_1 = (1/3)(11 C_A - 4N_fT_f)$ \cite{b1,casimir}, the asymptotic freedom
of the theory requires $N_f < N_{f,b1z}$, where $N_{f,b1z}=11C_A/(4T_f)$. 
(The subscript $b \ell z$ stands for ``$b_\ell$ zero''.)  Since
$\beta_{2\ell} = -[\alpha^2/(2\pi)](b_1+b_2 a)$, this function has an IR zero
at
\begin{equation}
\alpha_{IR,2\ell} = -\frac{4\pi b_1}{b_2} \ , 
\label{alfir_2loop}
\end{equation}
which is physical for $b_2 < 0$.  Now $b_2$ decreases linearly as a function of
$N_f$; for small $N_f$, $b_2 > 0$, but as $N_f$ increases through the value 
\begin{equation}
N_{f,b2z} = \frac{34 C_A^2}{4T_f(5C_A+3C_f)} \ ,
\label{nfb2z}
\end{equation}
$b_2$ reverses its sign and becomes negative.  Since $N_{f,b2z} < N_{f,b1z}$,
there is always an interval of $N_f < N_{f,b1z}$ for which $\beta$ has an IR
zero, namely the interval 
\begin{equation}
I: \quad N_{f,b2z} < N_f < N_{f,b1z} \ .
\label{nfinterval}
\end{equation}
If $R$ is the fundamental (fund.) representation of SU($N$), then 
\begin{equation}
I: \quad \frac{34N^3}{13N^2-3} < N_f < \frac{11N}{2} \ , \quad R = {\rm fund.} 
\label{nfinterval_fund} 
\end{equation}
For example, for $N=2$ and $N=3$, these intervals are $5.55 < N_f < 11$ and
$8.05 < N_f < 16.5$, respectively. As $N \to \infty$, the interval $I$ is
$2.62N < N_f < 5.5N$. Here and below, when an expression is given for $N_f$
that formally evaluates to a non-integral real value $\nu$, it is understood
implicitly that one infers an appropriate integral value of $N_f$ from it.

If $N_f \in I$ is near to $N_{f,b1z}$, so that $\alpha_{IR,2\ell}$ is small,
then the theory evolves from the UV to the IR without any spontaneous chiral
symmetry breaking. As $N_f$ decreases in $I$, $\alpha_{IR,2\ell}$ increases
and, as $N_f$ decreases through a critical value denoted $N_{f,cr}$, the IR
zero of $\beta$ increases through $\alpha_{cr}$, so that the UV to IR evolution
leads to S$\chi$SB.  For $N_f$ near lower end of $I$, $b_2 \to 0$ and
$\alpha_{IR,2\ell}$ is too large for the 2-loop calculation to be reliable.

\section{Calculations of Higher-Loop Corrections to UV to IR Evolution}

In this section we discuss our calculations of higher-loop corrections to the
UV to IR evolution of a gauge theory and, in particular, higher-loop
calculations of the IR zero of the beta function.  We first recall our 3-loop
calculation of the IR zero of $\beta$ in the $\overline{MS}$ scheme.  The
3-loop coefficient in $\beta$, $b_3$, is a quadratic function of $N_f$ and
vanishes, with sign reversal, at two values of $N_f$, denoted $N_{f,b3z,1}$ and
$N_{f,b3z,2}$. Now $b_3$ is positive for small $N_f$ and vanishes first at
$N_{f,b3z,1}$, which is smaller than $N_{f,b2z}$, the left endpoint of the
interval $I$. Furthermore, $N_{f,b3z,2} > N_{f,b1z}$, the right endpoint of
$I$. For example, for $N=2$, $N_{f,b3z,1}=3.99 < N_{f,b2z}=5.55$ and
$N_{f,b3z,2}=27.6 > N_{f,b1z}=11$, while for $N=3$, $N_{f,b3z,1}=5.84 <
N_{f,b2z}=8.05$ and $N_{f,b3z,2}=40.6 > N_{f,b1z}=16.5$.  Hence, $b_3 < 0$ for
$N_f \in I$, the interval of interest for the IR zero of $\beta$.  At this
3-loop level, $\beta_{3\ell} = -[\alpha^2/(2\pi)](b_1 + b_2 a + b_3 a^2)$, so
$\beta$ vanishes away from $\alpha=0$ at two values.  In terms of $\alpha$,
these are 
\begin{equation}
\alpha = \frac{2\pi}{b_3}\Big ( -b_2 \pm \sqrt{ b_2^2 - 4 b_1 b_3} \ \Big )
\end{equation}
Since $b_2 < 0$ in $I$ and the $\overline{MS}$ $b_3 < 0$ in $I$ also, this can
be expressed in terms of positive quantities as
\begin{equation}
\alpha=\frac{2\pi}{|b_3|}\Big ( -|b_2| \mp \sqrt{ b_2^2 + 4 b_1|b_3|} \ \Big )
\ . 
\end{equation}
One of these solutions is negative and hence is unphysical; the other is
manifestly positive, and is $\alpha_{IR,3\ell}$.  Using this result, we proved
in \cite{bvh} that in the $\overline{MS}$ scheme, 
$\alpha_{IR,3\ell} < \alpha_{IR,2\ell}$. 

A natural question that arises from the analysis in \cite{bvh} is how general
this inequality is and, specifically, whether it also holds for other
schemes. We address and answer this question here \cite{bc}.  To do this,
we observe that if a scheme had $b_3 > 0$ in $I$, then, since $b_2 \to 0$ at
the lower end of $I$, it would follow that $b_2^2-4b_1b_3 < 0$, so this scheme
would not have a physical (real, positive) $\alpha_{IR,3\ell}$ in this region.
Since the existence of the IR zero in $\beta_{2\ell}$ is a scheme-independent
property, one may require that an acceptable scheme should preserve this to
higher-loop order.  To satisfy this requirement, by the argument above, it
should have $b_3 < 0$ for $N_f \in I$, as is the case in the $\overline{MS}$
scheme.  We now prove that in all such schemes,
\begin{equation}
\alpha_{IR,3\ell} < \alpha_{IR,2\ell} \ . 
\label{alfir_23loop_ineq}
\end{equation}
To prove this, we consider the difference
\begin{equation}
\alpha_{IR,2\ell} - \alpha_{IR,3\ell} = \frac{2\pi}{|b_2 b_3|}
\bigg [ 2b_1 |b_3| + b_2^2 -|b_2|\sqrt{b_2^2+4b_1|b_3|} \ \bigg ] \ .
\label{alfir_2loop_minus_alfir_3loop}
\end{equation}
The expression in square brackets is positive if and only if
\begin{equation}
(2b_1 |b_3|+b_2^2)^2-b_2^2(b_2^2+4b_1|b_3|) > 0 \ .
\end{equation}
This difference is equal to the nonnegative quantity $(2b_1b_3)^2$, which
proves the inequality.  Note that, since $b_1 >0$ from the asymptotic
freedom property, this difference vanishes if and only if $b_3=0$, in which
case $\alpha_{IR,3\ell}=\alpha_{IR,2\ell}$.  This proof holds for general 
$G$, $R$, and $N_f \in I$. 

As noted above, $\alpha_{IR,2\ell}$ is a monotonically decreasing function of
$N_f \in I$.  With $b_3 < 0$ for $N_f \in I$, this monotonicity property is
also true of $\alpha_{IR,3\ell}$. As $N_f$ increases from $N_{f,b2z}$ to
$N_{f,b1z}$ in the interval $I$, $\alpha_{IR,3\ell}$ decreases from
\begin{equation}
\alpha_{IR,3\ell} = 4\pi \sqrt{ \frac{b_1}{|b_3|}} \quad {\rm at} \ N_f =
N_{f,b2z} 
\label{alf_3loop_at_nfb2z}
\end{equation}
to zero at the upper end of the interval $I$, vanishing like
\begin{equation}
\alpha_{IR,3\ell} = \frac{4\pi b_1}{|b_2|} \bigg [1 - \frac{|b_3|b_1}{|b_2|^2}
+ O( b_1^2 ) \ \bigg ] \quad {\rm as} \ \ N_f \nearrow N_{f,b1z} \ .
\label{alfir_3loop_upperend}
\end{equation}

For the 4-loop analysis of $\beta$, we use $b_4$, which is a cubic polynomial
in $N_f$. This coefficient is positive for $N_f \in I$ for $N=2,3$ but is
negative in part of $I$ for higher $N$. The 4-loop $\beta$ function is
$\beta_{4\ell} = -[\alpha^2/(2\pi)](b_1+b_2a+b_3a^2+b_4a^3)$, so $\beta$ has
three zeros away from the origin.  We determine the smallest positive real zero
as $\alpha_{IR,4\ell}$.  

In addition to the inequality (\ref{alfir_23loop_ineq}), we find the following
general results: (i) going from the 3-loop to 4-loop level, there is a slight
change in the value of the IR zero, but this change is smaller than the
decrease from the 2-loop to 3-loop level, so $\alpha_{IR,4\ell} <
\alpha_{IR,2\ell}$; and (ii) fractional changes in the value of the IR zero of
$\beta$ decrease in magnitude as $N_f$ increases toward $N_{f,b1z}$, and all of
the values of $\alpha_{IR,n\ell} \to 0$.  Our finding that the fractional
change in the location of the IR zero of $\beta$ is reduced at higher-loop
order agrees with the general expectation that calculating a quantity to higher
order in perturbation theory should give a more stable and accurate result.
Since $\alpha_{cr} \sim O(1)$, the decrease in $\alpha_{IR}$ at higher-loop
order, together with the property that $\alpha_{IR}$ increases as $N_f$
decreases, means that one must go to smaller $N_f$ for $\alpha_{IR,n\ell}$ to
grow to a given size for the $n=3$ and $n=4$ loop level as compared with the
$n=2$ loop level. This suggests that the actual lower boundary of the
IR-conformal phase could lie somewhat below an estimate obtained setting
$\alpha_{IR,2\ell}=\alpha_{cr}$. 

Some numerical values of $\alpha_{IR,n\ell}$ at the 2-loop, 3-loop, and
4-loop level for fermions in the fundamental representation, 
$N_f \in I$, and the illustrative groups $G={\rm SU}(2)$ and $G={\rm SU}(3)$
from \cite{bvh} are given in Table 1.  

\begin{table}
\tbl{Values of $\alpha_{IR,n\ell}$ for the beta function of
an SU($N$) gauge theory with $N_f$ fermions in the fundamental
representation, for $N=2, \ 3$ and $N_f \in I$.}
{\begin{tabular}{|c|c|c|c|c|c|}
\toprule
$N$ & $N_f$& $\alpha_{IR,2\ell}$ & $\alpha_{IR,3\ell}$ &$\alpha_{IR,4\ell}$ 
\\\colrule
 2  &  7  &  2.83   & 1.05   & 1.21     \\
 2  &  8  &  1.26   & 0.688  & 0.760    \\
 2  &  9  &  0.595  & 0.418  & 0.444    \\
 2  & 10  &  0.231  & 0.196  & 0.200    \\
 \hline
 3  & 10  &  2.21   & 0.764  & 0.815    \\
 3  & 11  &  1.23   & 0.578  & 0.626    \\
 3  & 12  &  0.754  & 0.435  & 0.470    \\
 3  & 13  &  0.468  & 0.317  & 0.337    \\
 3  & 14  &  0.278  & 0.215  & 0.224    \\
 3  & 15  &  0.143  & 0.123  & 0.126    \\
 3  & 16  &  0.0416 & 0.0397 & 0.0398   \\\botrule
\end{tabular}}
\bigskip
\label{alfirvalues}
\end{table}

Corresponding higher-loop calculations were carried out in \cite{bvh} for
SU($N$) gauge theories with $N_f$ fermions in the adjoint, symmetric and
antisymmetric rank-2 tensor representations.  The general result
$\alpha_{IR,3\ell} < \alpha_{IR,2\ell}$ continues to apply. The difference
$\alpha_{IR,4\ell}-\alpha_{IR,3\ell}$ tends to be relatively small, but can
have either sign.  For example, for $R=$ adjoint, $N_{f,b1z}=11/4$ and
$N_{f,b2z}=17/16$ (independent of $N$), so the interval $I$ where $\beta$ has
an IR zero, viz., $N_{f,b2z} < N_f < N_{f,b1z}$, is $1.06 < N_f < 2.75$. This
includes only one physical, integral value, $N_f=2$. For this value of $N_f$
and $N=2$, one has $\alpha_{IR,2\ell}=0.628$, $\alpha_{IR,3\ell}=0.459$,
$\alpha_{IR,4\ell}=0.450$, while for $N=3$, $\alpha_{IR,2\ell}=0.419$,
$\alpha_{IR,3\ell}=0.306$, $\alpha_{IR,4\ell}=0.308$.

\section{Anomalous Dimension of Fermion Bilinear}

The anomalous dimension $\gamma_m \equiv \gamma$ for the fermion bilinear
operator has the series expansion 
\begin{equation}
\gamma = \sum_{\ell=1}^\infty c_\ell a^\ell
   = \sum_{\ell=1}^\infty \bar c_\ell \alpha^\ell \ , 
\end{equation}
where $\bar c_\ell = c_\ell/(4\pi)^\ell$ is the $\ell$-loop coefficient. 
The one-loop coefficient $c_1$ is scheme-independent, while the 
$c_\ell$ with $\ell \ge 2$ are scheme-dependent. The $c_\ell$ have been
calculated up to 4-loop level in the $\overline{MS}$ scheme \cite{gamma4}.  
The first two coefficients are $c_1 = 6C_f$ and 
$c_2=2C_f[(3/2)C_f + (97/6)C_A-(10/3)T_fN_f]$. 

It is of interest to calculate $\gamma$ at the exact IRFP in the IR-conformal
phase and at the approximate IRFP in the phase with spontaneous chiral symmetry
breaking. We denote $\gamma$ calculated to $n$-loop ($n\ell$) level as
$\gamma_{n\ell}$ and, evaluated at the IR zero of
$\beta_{n\ell}$, as $\gamma_{IR,n\ell} \equiv
\gamma_{n\ell}(\alpha=\alpha_{IR,n\ell})$.  In the IR-conformal phase, an exact
calculation of $\gamma$ evaluated at the IRFP would be an exact
(scheme-independent) property of the theory, but in the broken phase, just as
the IR zero of $\beta$ is only an approximate IRFP, so also, $\gamma$ is only
approximate, describing the running of $\bar\psi\psi$ and the dynamically
generated fermion mass near the zero of $\beta$, according to
\begin{equation}
\Sigma(k) \sim \Lambda \Big ( \frac{\Lambda}{k} \Big )^{2-\gamma}
\label{sigmak}
\end{equation}
for Euclidean momenta $k >> \Lambda$.  This, in
turn, affects SM fermion masses in DEWSB theories \cite{wtc}. In both the
conformal and the chirally broken IR phases, the upper bound $\gamma < 2$ 
holds. 

\begin{figure}
  \begin{center}
    \includegraphics[height=8cm]{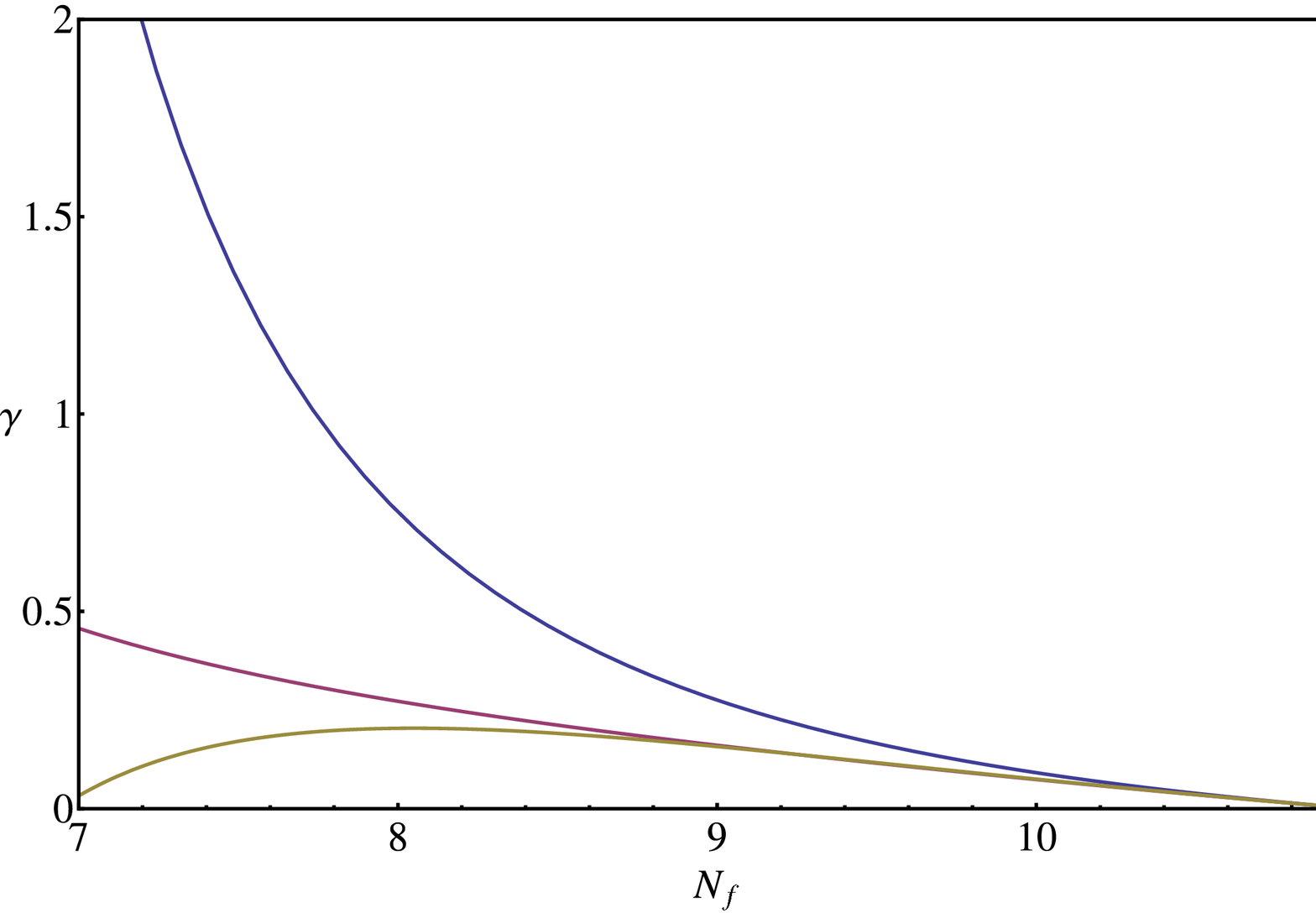}
  \end{center}
\caption{$\gamma_{IR}$ for SU(2) and fermions in the fundamental
  representation. From top to bottom, the curves show 
 $\gamma_{IR,2\ell}$, $\gamma_{IR,3\ell}$, and 
$\gamma_{IR,4\ell}$.}
\label{gam_fund_nc2}
\end{figure}

\begin{figure}
  \begin{center}
    \includegraphics[height=8cm]{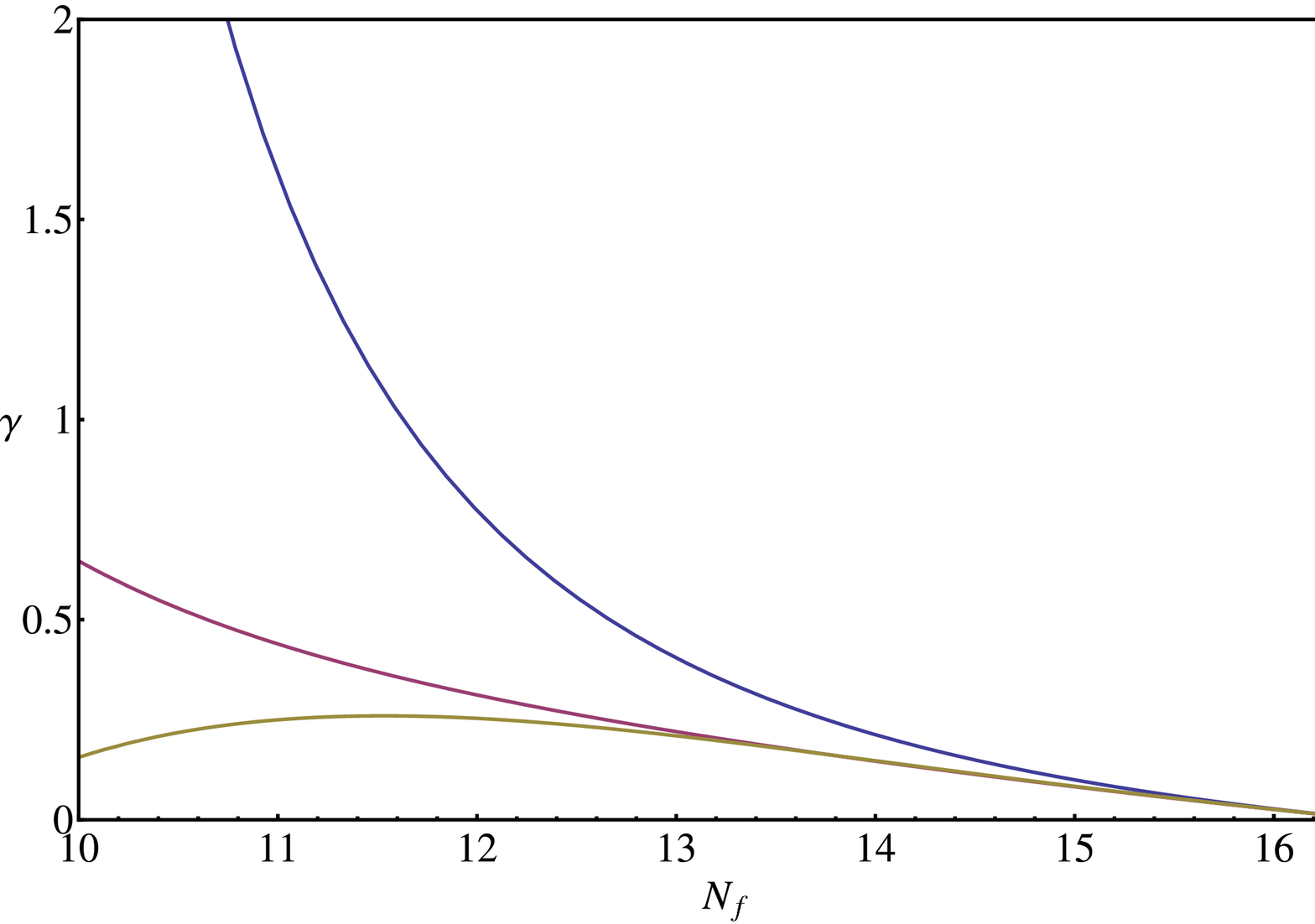}
  \end{center}
\caption{$\gamma_{IR}$ for SU(3) and fermions in the fundamental
  representation. From top to bottom, the curves show 
 $\gamma_{IR,2\ell}$, $\gamma_{IR,3\ell}$, and 
$\gamma_{IR,4\ell}$.}
\label{gam_fund_nc3}
\end{figure}

At the 2-loop level we calculate 
\begin{equation}
\gamma_{IR,2\ell} = \frac{C_f(11C_A-4T_fN_f)[455C_A^2+99C_AC_f 
+ (180C_f-248C_A)T_fN_f+80(T_fN_f)^2]}{12[-17C_A^2+2(5C_A+3C_f)T_fN_f]^2}
\end{equation}
Our analytic expressions for $\gamma_{IR,n\ell}$ at the 3-loop and 4-loop level
are more complicated \cite{bvh}.  Illustrative numerical values of
$\gamma_{IR,n\ell}$ at the 2-, 3-, and 4-loop level are given below in Table
\ref{gammavalues} and Figs. \ref{gam_fund_nc2} and \ref{gam_fund_nc3} for
$R=fund.$, and the illustrative values $N=2, \ 3$.  (Values in parentheses
violate the upper bound $\gamma < 2$ and reflect the inadequacy of perturbation
theory if $\alpha$ is too large.)

\begin{table}
\tbl{Values of $\gamma_{IR,n\ell}$ for an SU($N$) gauge theory with 
$N_f$ fermions in the fundamental representation, for $N=2, \ 3$ and 
$N_f \in I$.}
{\begin{tabular}{|c|c|c|c|c|}
\toprule
$N$ & $N_f$& $\gamma_{IR,2\ell}$ & $\gamma_{IR,3\ell}$ & $\gamma_{IR,4\ell}$
\\\colrule
 2  &  7  & (2.67) & 0.457  & 0.0325  \\
 2  &  8  & 0.752  & 0.272  & 0.204   \\
 2  &  9  & 0.275  & 0.161  & 0.157   \\
 2  & 10  & 0.0910 & 0.0738 & 0.0748  \\
 \hline
 3  & 10  & (4.19) & 0.647  & 0.156   \\
 3  & 11  & 1.61   & 0.439  & 0.250   \\
 3  & 12  & 0.773  & 0.312  & 0.253   \\
 3  & 13  & 0.404  & 0.220  & 0.210   \\
 3  & 14  & 0.212  & 0.146  & 0.147   \\
 3  & 15  & 0.0997 & 0.0826 & 0.0836  \\
 3  & 16  & 0.0272 & 0.0258 & 0.0259  \\\botrule
\end{tabular}}
\bigskip
\label{gammavalues}
\end{table}


We have also performed these calculations for $G={\rm SU}(N)$ and higher
fermion representations $R$.  In general, we find that, for a given $N$, $R$,
and $N_f$, the values of $\gamma_{IR,n\ell}$ calculated to 3-loop and 4-loop
order are smaller than the 2-loop value. The value of these higher-loop
calculations is evident from the figures.  A necessary condition for a
perturbative calculation to be reliable is that higher-order contributions do
not modify the result too much.  One sees from the tables and figures that,
especially for smaller $N_f$, there is a substantial decrease in
$\alpha_{IR,n\ell}$ and $\gamma_{IR,n\ell}$ when one goes from 2-loop to 3-loop
order, but for a reasonable range of $N_f$, the 3-loop and 4-loop results are
close to each other.  Thus, our higher-loop calculations of $\alpha_{IR}$ and
$\gamma$ allow us to probe the theory reliably down to smaller values of $N_f$
and thus stronger couplings.

It is useful to give a comparison of our calculations with lattice
measurements.  The theory with SU(3), $R=fund.$, and $N_f=12$ has been the
subject of intensive lattice study, so we focus on this for the comparison.
For this theory we calculate (see Table \ref{gammavalues})
$\gamma_{IR,2\ell}=0.77$, $\gamma_{IR,3\ell}=0.31$, and
$\gamma_{IR,4\ell}=0.25$ (to two significant figures). Lattice results include
$\gamma=0.414 \pm 0.016$ \cite{lsdgamma}; $\gamma \sim 0.35$ \cite{degrand};
$0.2 \le \gamma \le 0.4$ \cite{kuti}; $\gamma =0.4 - 0.5$ \cite{latkmi}; and
$\gamma=0.27 \pm 0.03$ \cite{hasenfratz}.  Here the 2-loop value of $\gamma$ is
larger than, and the 3-loop and 4-loop values are closer to, these lattice
measurements.  This illustrates how higher-loop calculations of $\gamma$ can
improve agreement with lattice measurements. Note that there is not yet a
consensus among lattice groups as to whether this theory has an IR phase with
chiral symmetry or spontaneous chiral symmetry breaking. Similar comparisons
can be given for other values of $G$, $R$ and $N_f$.  In particular, for SU(3),
$R=fund.$, $N_f=10$, one group obtains $\gamma_{IR} \sim 1$ \cite{lsdnf10},
consistent with the idea that $\gamma_{IR} \sim 1$ at the lower end of the
IR-conformal phase. This region is difficult to probe perturbatively because of
the strongly coupled nature of the physics.

An interesting property of the values of $\alpha_{IR,n\ell}$ and
$\gamma_{IR,n\ell}$ in the case where $R=fund.$ is that these are similar in 
theories with different values of $N$ and $N_f$, provided that these theories
have the same or similar values of the ratio $N_f/N$.  This can be understood
as a result of a rapid approach to the 't Hooft-Veneziano limit 
$N \to \infty$, $N_f \to \infty$ with $N_f/N$ fixed \cite{bvh,bc,lnn}.  

\section{Supersymmetric Gauge Theory}

It is valuable to carry out a similar analysis in an asymptotically free ${\cal
N}=1$ supersymmetric gauge theory with vectorial chiral superfield content
$\Phi_i, \ \tilde \Phi_i$, $i=1,...,N_f$ in the $R, \ \bar R$ reps.,
respectively \cite{bfs}. An appeal of this analysis is that exact results on
the IR properties of the theory are known \cite{seiberg}. Thus, one can compare
results from higher-loop perturbative calculations with exact results, in
particular, for $N_{f,cr}$.  The coefficients of $\beta$ and $\gamma$ have been
calculated up to 3-loop order \cite{susyloops}. The one-loop coefficient is
$b_{1,s} = 3C_A - 2T_fN_f$. (Here and below, the subscript $s$, for
``supersymmetric'', is appended to distinguish these quantities from those for
the nonsupersymmetric theory). Asymptotic freedom requires that $N_f <
N_{f,b1z,s}$, where $N_{f,b1z,s} = 3C_A/(2T_f)$. 
The 2-loop coefficient of $\beta_s$ is $b_{2,s} =
6C_A^2-4T_fN_f(C_A+2C_f)$, which is positive for small $N_f$, but vanishes with
sign reversal as $N_f$ increases through the value
\begin{equation}
N_{f,b2z,s} = \frac{3C_A^2}{2T_f(C_A+2C_f)} \ ,
\label{nfb2z_susy}
\end{equation}
which is smaller than $N_{f,b1z,s}$.  Thus, for this
theory, there is again an interval $I_s$ in which $\beta_{2\ell}$ has an IR
zero, namely
\begin{equation}
I_s: \quad N_{f,b2z,s} < N_f < N_{f,b1z,s} \ .
\end{equation}
For $R=fund.$ this interval $I_s$ is
$3N^3/(2N^2-1) < N_f < 3N$.  In general, this IR zero of $\beta_{2\ell,s}$ 
occurs at
\begin{equation}
\alpha_{IR,2\ell,s} = \frac{2\pi(3C_A-2T_fN_f)}{2T_fN_f(C_A+2C_f)-3C_A^2} \ . 
\label{alfir_2loop_susy}
\end{equation}

The 3-loop coefficient $b_{3,s}$ is positive for small $N_f$ and vanishes
at two values of $N_f$, denoted $N_{f,b3z,1,s}$ and $N_{f,b3z,2,s}$. As before,
we find that $N_{f,b3z,1,s} < N_{f,b2z,s}$ and
$N_{f,b3z,2,s} > N_{f,b1z,s}$, so $b_{3,s} < 0$ for $N_f \in I_s$.
Since $b_{3,s} < 0$ for $N_f \in I$, we find, by the same type of proof as
given above, that for any $G$, $R$, and $N_f \in I_s$ 
\begin{equation}
\alpha_{IR,3\ell,s} < \alpha_{IR,2\ell,s} \ . 
\end{equation}
For fixed $N$, we find that $\alpha_{IR,n\ell}$ increases monotonically with
decreasing $N_f$ at both the 2-loop and 3-loop level.

Next, we analyze the anomalous dimension $\gamma$ of the superfield operator
product $\Phi \tilde \Phi$ containing the term $\theta\theta\psi \tilde \psi$.
In a conformally invariant $d$-dimensional field theory (whether supersymmetric
or not), unitarity yields a lower bound on the dimension $D_{\cal O}$ of a
spin-0 operator ${\cal O}$ (other than the identity), namely, $D_{\cal O} \ge
(d-2)/2$, where $d=$ spacetime dim.; so $D_{\cal O} \ge 1$ here
\cite{gbound}. In the nonsupersymmetric theory, with ${\rm dim}(\bar \psi \psi)
= 3-\gamma$, this constraint is $D_{\bar\psi\psi} = 3-\gamma > 1$, so $\gamma <
2$.  In the supersymmetric theory, with ${\rm dim}(\theta)=-1/2$ and ${\rm
dim}(\psi \tilde\psi) = 3-\gamma$, the constraint is $D_{\Phi \tilde \Phi} =
-1+3-\gamma > 1$, so $\gamma < 1$.

At the 2-loop level, we find
\begin{equation}
\gamma_{IR,2\ell,s} = \frac{C_f(3C_A-2T_fN_f)(2T_fN_f-C_A)(2T_fN_f-3C_A+6C_f)}
{[2(C_A+2C_f)T_fN_f-3C_A^2]^2} \ . 
\end{equation}
We have also calculated $\gamma_{IR,3\ell}$ and find that, as in the
nonsupersymmetric case, $\gamma_{IR,3\ell} < \gamma_{IR,2\ell}$.  Let us focus
on the case $R=fund.$, for which $N_{f,b1z,s}=3N$ and $N_{f,cr,s}=(3/2)N$. One
perturbative estimate of $N_{f,cr}$ can be obtained by assuming that the upper
bound $\gamma \le 1$ is saturated as $N_f \searrow N_{f,cr,s}$.  Solving this,
we obtain that for $N=2$, $N_{f,cr,s,est.}=4.24$, a factor of 1.41 larger than
exact $N_{f,cr,s}=3$, while for $N=3$ $N_{f,cr,s,est.}=6.15$ a factor of 1.37
larger than exact $N_{f,cr,s}=4.5$.  As $N \to \infty$, this procedure yields
$N_{f,cr,s,est.} \to 2N$, a factor of (4/3) larger than the exact result,
$N_{f,cr,s}=(3/2)N$.

This comparison for the ${\cal N}=1$ supersymmetric gauge theory suggests that
the perturbative calculation slightly overestimates the value of $N_{f,cr}$, 
i.e., slightly underestimates the size of the IR-conformal phase, similar to
what we found for the nonsupersymmetric theory. 

\section{Scheme-Dependence in Calculation of IR Fixed Point}

Since the coefficients in $\beta$ at the level of $n \ge 3$ loops are 
scheme-dependent, so is the resultant value of $\alpha_{IR,n\ell}$. 
It is important to assess quantitatively the uncertainty in the analysis of the
UV to IR evolution due to this scheme dependence.
A way to do this is to perform scheme transformations and determine how much of
a change there is in $\alpha_{IR,n\ell}$ \cite{sch}. 

A scheme transformation (ST) is a map between $\alpha$ and $\alpha'$ or
equivalently, $a$ and $a'$, where $a=\alpha/(4\pi)$, which can be expressed as
\begin{equation}
a = a'f(a')
\end{equation}
with $f(0)=1$ to keep the UV properties unchanged. We write 
\begin{equation}
f(a') = 1 + \sum_{s=1}^{s_{max}} k_s (a')^s =
        1 + \sum_{s=1}^{s_{max}} \bar k_s (\alpha')^s \ ,
\label{faprime}
\end{equation}
where the $k_s$ are constants, $\bar k_s = k_s/(4\pi)^s$, and $s_{max}$ may be
finite or infinite. Hence, the Jacobian
\begin{equation}
J = \frac{da}{da'} = \frac{d\alpha}{d\alpha'} 
\label{j}
\end{equation}
satisfies $J=1$ at $a=a'=0$. We have
\begin{equation}
\beta_{\alpha'} \equiv \frac{d\alpha'}{dt} = \frac{d\alpha'}{d\alpha} \,
\frac{d\alpha}{dt} = J^{-1} \, \beta_{\alpha} \ .
\label{betaap}
\end{equation}
This has the expansion
\begin{equation}
\beta_{\alpha'} = -2\alpha' \sum_{\ell=1}^\infty b_\ell' (a')^\ell =
-2\alpha' \sum_{\ell=1}^\infty \bar b_\ell' (\alpha')^\ell \ ,
\label{betaprime}
\end{equation}
where $\bar b'_\ell = b'_\ell/(4\pi)^\ell$.

Using these two equivalent expressions for $\beta_{\alpha'}$, one can
solve for the $b_\ell'$ in terms of the $b_\ell$ and $k_s$.
This yields the well-known result that $b_1' = b_1$ and $b_2' = b_2$.
To assess the scheme-dependence of an IRFP, we have calculated the relations
between the $b'_\ell$ and $b_\ell$ for higher $\ell$ values \cite{sch}. For
example, for $\ell=3, \ 4, \ 5$, we obtain 
\begin{equation}
b_3' = b_3 + k_1b_2+(k_1^2-k_2)b_1 \ ,
\end{equation}
\begin{equation}
b_4' = b_4 + 2k_1b_3+k_1^2b_2+(-2k_1^3+4k_1k_2-2k_3)b_1
\end{equation}
\begin{eqnarray}
b_5' & = & b_5+3k_1b_4+(2k_1^2+k_2)b_3+(-k_1^3+3k_1k_2-k_3)b_2 \cr\cr
     & + & (4k_1^4-11k_1^2k_2+6k_1k_3+4k_2^2-3k_4)b_1 \ . 
\end{eqnarray}

To be physically acceptable, a ST must satisfy several conditions, $C_i$:
\begin{itemize}

\item

$C_1$: the ST must map a real positive $\alpha$ to a real
positive $\alpha'$, since a map taking $\alpha > 0$ to $\alpha'=0$ would be
singular, and a map taking $\alpha > 0$ to a negative or complex $\alpha'$
would violate the unitarity of the theory.

\item

$C_2$: the ST should not map a moderate value of $\alpha$, for which
perturbative calculations may be reliable, to an excessively large value of
$\alpha'$ where perturbative calculations are inapplicable.

\item

$C_3$: \ $J$ should not vanish in the region of $\alpha$ and
$\alpha'$ of interest, or else there would be a pole in the relation between
$\beta_\alpha$ and $\beta_{\alpha'}$.

\item

$C_4$: The existence of an IR zero of $\beta$ is a scheme-independent property,
depending (in an asymptotically free theory) only on the condition that $b_2 <
0$.  Hence, a ST should satisfy the condition that $\beta_\alpha$ has an IR
zero if and only if $\beta_{\alpha'}$ has an IR zero.

\end{itemize}

These four conditions can always be satisfied by scheme transformations near a
UV fixed point, and hence in applications to perturbative QCD calculation,
since $\alpha$ is small, and one can choose the $k_s$ to be small also, so
$\alpha' \simeq \alpha$. However, these conditions C1-C4 are not automatically
satisfied, and are a significant constraint, on a scheme transformation applied
in the vicinity of an IRFP, where $\alpha$ may be O(1). For example, consider
the scheme transformation
\begin{equation}
\alpha = \tanh(\alpha')
\end{equation}
with inverse
\begin{equation}
\alpha' = \frac{1}{2} \ln \Big ( \frac{1+\alpha}{1-\alpha} \Big ) \ . 
\end{equation}
If $\alpha << 1$, as at a UVFP, this is acceptable, but as
$\alpha$ approaches 1 from below it maps a moderate value of $\alpha$ 
to an arbitrarily large $\alpha'$ and hence fails condition C2, and 
if $\alpha$ exceeds 1, even if by a small amount, then it fails 
conditions C1 and C4, since it maps a real positive $\alpha$ to a complex 
$\alpha'$. 

We have studied the scheme dependence of the IR zero of $\beta$ in \cite{sch}
using several scheme transformations; e.g., the ST (depending on a parameter
$r$)
\begin{equation}
S_{sh,r}: \quad a=\frac{\sinh(ra')}{r}
\end{equation}
Since $\sinh(ra')/r$ is an even fn. of $r$, we take $r > 0$ with no
loss of generality. This transformation has the inverse
\begin{equation}
a' = \frac{1}{r} \, \ln \Big [ ra + \sqrt{1+ (ra)^2} \ \Big ]
\end{equation}
and the Jacobian $J = \cosh(ra')$. For this ST,
\begin{equation}
f(a') = \frac{\sinh(r a')}{r a'} \ .
\end{equation}
This $f(a')$ has a series expansion with $k_s=0$ for odd $s$ and, for even $s$,
\begin{equation}
k_2 = \frac{r^2}{6} \ , \quad k_4 = \frac{r^4}{120} \ , \quad 
k_6 = \frac{r^6}{5040} \ , \quad k_8 = \frac{r^8}{362880} \ ,
\end{equation}
etc. for higher $s$. Substituting these results for $k_s$ into the equations
for the $b_\ell'$, we obtain
\begin{equation}
b'_3 = b_3 - \frac{r^2 b_1}{6} \ , \quad b'_4 = b_4 \ , 
\end{equation}
\begin{equation}
b'_5 = b_5 + \frac{r^2b_3}{6} + \frac{31r^4b_1}{360} \ , 
\end{equation}
and so forth for higher $\ell$.

We apply this $S_{sh_r}$ ST to the $\beta$ function in the
$\overline{MS}$ scheme, calculated up to $\ell=4$ loop level.
For $N_f \in I$ where $\beta_{2\ell}$ has an IR zero, we then calculate the 
resultant IR zeros in $\beta_{\alpha'}$ at the 3-loop and 4-loop order and 
compare the values with those in the $\overline{MS}$ scheme.
We list some numerical results for illustrative values of $r$ and for $N=2, \
3$ below.  We denote the IR zero of $\beta_{\alpha'}$ at the $n$-loop level as
$\alpha'_{IR,n\ell} \equiv \alpha'_{IR,n\ell,r}$.
For example, for $N=3$, $N_f=10$, $\alpha_{IR,2\ell}=2.21$, and:
\begin{equation}
\alpha_{IR,3\ell,\overline{MS}}=0.764, \quad
\alpha'_{IR,3\ell,r=3}=0.762, \quad \alpha'_{IR,3\ell,r=6}=0.754,
\end{equation}
\begin{equation}
\alpha'_{IR,3\ell,r=9}=0.742, \quad \alpha'_{IR,3\ell,r=4\pi}=0.723
\end{equation}
\begin{equation}
\alpha_{IR,4\ell,\overline{MS}}=0.815, \quad
\alpha'_{IR,4\ell,r=3}=0.812, \quad \alpha'_{IR,4\ell,r=6}=0.802,
\end{equation}
\begin{equation}
\alpha'_{IR,4\ell,r=9}=0.786, \quad \alpha'_{IR,4\ell,r=4\pi}=0.762 \ . 
\end{equation}
In general, the effect of scheme dependence tends to be reduced (i) for a given
$N$ and $N_f$, as one calculates to higher-loop order, and (ii) for a given
$N$, as $N_f \to N_{f,b1z}$, so that the value of $\alpha_{IR} \to 0$.  These
results provide a quantitative measure of scheme dependence of the location of
an IR zero of $\beta$.

Since the $\beta$ function coefficients $b_\ell$ with $\ell \ge 3$ are
scheme-dependent, there should exist a ST that renders these coefficients equal
to zero (i.e., maps to the 't Hooft scheme).  We constructed an explicit 
ST that can does this at a UVFP. This necessarily has $s_{max}=\infty$. 
For simplicity, we set $k_1=0$.  Our solutions for the first few 
$k_s$ are 
\begin{equation}
k_2 = \frac{b_3}{b_1} \ , \quad k_3 = \frac{b_4}{2b_1} \ , 
\label{k23tooftsol}
\end{equation}
\begin{equation}
k_4 = \frac{b_5}{3b_1} - \frac{b_2b_4}{6b_1^2} + \frac{5b_3^2}{3b_1^2} \ , 
\label{k4thooftsol}
\end{equation}
\begin{equation}
k_5 = \frac{b_6}{4b_1} - \frac{b_2b_5}{6b_1^2} + \frac{2b_3b_4}{b_1^2}
+ \frac{b_2^2b_4}{12b_1^3} - \frac{b_2b_3^2}{12b_1^3} \ , 
\label{k5thooftsol}
\end{equation}
and so forth for higher $s$.

\section{Application to Models of Dynamical Electroweak Symmetry Breaking} 

Models with dynamical EWSB have been of interest as one way to avoid the
hierarchy (fine-tuning) problem with the Standard Model, while supersymmetry
provides another way to do this.  DEWSB models use an asymptotically free
vectorial gauge interaction with a certain set of fermions subject to this
gauge interaction, which becomes strong at the TeV scale, producing bilinear
condensates of these fermions. These models involve further ingredients to give
masses to SM fermions.  The running mass $m_{f_i}(p)$ of a SM fermion of
generation $i$ is constant up to the scale where it is dynamically generated,
and has a power-law decay above this scale \cite{sml}.  Quasiconformal behavior
in these models resulting from an approximate IRFP \cite{wtc} is important as a
means to produce a substantial $\gamma_m$ and hence enhance SM fermion masses.
Since the quasiconformal property depends on having an approximate IRFP
$\alpha_{IR}$ slightly greater than $\alpha_{cr}$, the higher-loop calculations
of the UV to IR evolution described here are valuable not only for their
intrinsic field-theoretic interest, but also as applied to the construction of
such models. Studies of reasonably UV-complete DEWSB models showed that
approximate residual generational symmetries suppress FCNC effects
\cite{ckm,kt}, but it is challenging to reproduce all features of SM fermion
masses, such as $m_t >> m_b$, etc.

Concerning dilatons resulting from quasiconformal DEWSB models, we note that
such models need not have any color-nonsinglet fermions subject to the new
gauge interaction. The minimal SM-nonsinglet fermion content consists of one
SU(2)$_L$ doublet with corresponding right-handed SU(2)$_L$ singlets, all of
which are color-singlets. DEWSB models of this type can exhibit quasiconformal
behavior \cite{ts}.  Hence, in contrast to one-family DEWSB models, in these
minimal models, the resultant dilaton would be comprised purely of
color-singlet constituents. This difference is recognized in some of the
phenomenological papers on dilatons.  The boson with mass $\simeq 125$ GeV
discovered at the LHC by ATLAS and CMS is consistent with being the SM
Higgs. An important question is to determine how well it can, alternatively, be
modelled as a dilaton resulting from a quasiconformal DEWSB model. Further
experimental and theoretical work should settle this question.

\section{Conclusions}

In conclusion, understanding the UV to IR evolution of an asymptotically free
gauge theory and the nature of the IR behavior is of fundamental
field-theoretic interest.  Our higher-loop calculations give new information on
this UV to IR flow and on the determination of $\alpha_{IR,n\ell}$ and
$\gamma_{IR,n\ell}$.  It is valuable to compare our higher-order calculations
of $\gamma_{IR,n\ell}$ with lattice measurements.  The higher-loop study of the
UV to IR flow for supersymmetric gauge theories yields further insights.  We
have carried out a study of scheme-dependence in higher-loop calculations,
noting that scheme transformations are subject to constraints that are easily
satisfied at a UVFP but are quite restrictive at IRFP.  In addition to their
intrinsic interest, quasiconformal gauge theories are relevant to models of
dynamical EWSB, plausibly yielding a light pseudo-Nambu-Goldstone boson, the
dilaton.

\bigskip
\bigskip

I thank K. Yamawaki for the invitation to speak at SCGT12 and the excellent
organization of the conference. I thank T. Ryttov for collaboration on
\cite{bvh,bfs,sch}. This talk was given on Dec. 5, 2012 at Nagoya University,
during a sabbatical at Yale University, and I thank T. Appelquist and the
theory group at Yale for warm hospitality during the sabbatical. This research
was partially supported by the grant NSF-PHY-09-69739.


\begin{thebibliography}{9}
\bibitem{casimir}
%
Here $C_A \equiv C_2(G)$ is the quadratic Casimir invariant for the adjoint 
representation, $C_f \equiv C_2(R)$, and $T_f$ is the trace invariant.
\end{thebibliography}
\end{document}